\documentclass[]{article}

\usepackage{color}
\usepackage{graphicx}
\usepackage{mathptmx}

\begin{document}
	
{\bf \Large Noise induces continuous and noncontinuous transitions in neuronal
		interspike intervals range}\\
	
	P R Protachevicz\textsuperscript{1,2}, M S
		Santos\textsuperscript{2}, E G Seifert\textsuperscript{3}, E C
		Gabrick\textsuperscript{1}, F S Borges\textsuperscript{4}, R R
		Borges\textsuperscript{5}, J Trobia\textsuperscript{1,6}, J D Szezech
		Jr\textsuperscript{1,6}, K C Iarosz\textsuperscript{2,5,7,*}, I L
		Caldas\textsuperscript{2}, C G Antonopoulos\textsuperscript{8}, Y
		Xu\textsuperscript{9}, R L Viana\textsuperscript{10}, A M
		Batista\textsuperscript{1,2,6}\\
	
{\footnotesize	\textsuperscript{1}Graduate Program in Science - Physics, State
		University of Ponta Grossa, 84030-900, Ponta Grossa, PR, Brazil\\
	\textsuperscript{2}Institute of Physics, University of S\~ao Paulo,
		05508-900, S\~ao Paulo, SP, Brazil\\
	\textsuperscript{3}Department of Physics, State University of Ponta
		Grossa, 84030-900, Ponta Grossa, PR, Brazil\\
	\textsuperscript{4}Center for Mathematics, Computation, and
		Cognition, Federal University of ABC, 09606-045, S\~ao Bernardo do Campo, SP,
		Brazil\\
	\textsuperscript{5}Department of Mathematics, Federal University of
		Technology Paran\'a, 84016-210, Ponta Grossa, PR, Brazil\\
	\textsuperscript{6}Department of Mathematics and Statistics, State
		University of Ponta Grossa, 84030-900, Ponta Grossa, PR, Brazil\\
\textsuperscript{7}Faculdade de Tel\^emaco Borba, FATEB, 84266-010,
		Tel\^emaco Borba, PR, Brazil\\
	\textsuperscript{8}Department of Mathematical Sciences, University
		of Essex, CO4 3SQ, Wivenhoe Park, UK\\
	\textsuperscript{9}Department of Applied Mathematics, Northwestern
		Polytechnical University, 710072, Xi'an, China\\
	\textsuperscript{10}Department of Physics, Federal University of
		Paran\'a, 82590-300, Curitiba, PR, Brazil\\}
		
		Corresponding author: protachevicz@gmail.com.
		
		\begin{abstract}  
			Noise appears in the brain due to various sources, such as ionic channel
			fluctuations and synaptic events. They affect the activities of the brain and
			influence neuron action potentials. Stochastic differential equations have been
			used to model firing patterns of neurons subject to noise. In this work, we
			consider perturbing noise in the adaptive exponential integrate-and-fire (AEIF)
			neuron. The AEIF is a two-dimensional model that describes different neuronal
			firing patterns by varying its parameters. Noise is added in the equation
			related to the membrane potential. We show that a noise current can induce
			continuous and noncontinuous transitions in neuronal interspike intervals.
			Moreover, we show that the noncontinuous transition occurs mainly for
			parameters close to the border between tonic spiking and burst activities of
			the neuron without noise.
		\end{abstract}
		
		{noise, adaptive exponential integrate-and-fire, neuronal activities}

	
	\section{Introduction}
	
	Neuronal activities play an important role in brain functions. In the late
	1800s, Caton \cite{caton75} and Beck \cite{coenen13} recor\-ded electrical brain
	activities in various animal species by means of electrodes. Hans
	\cite{vaque99} discovered the electroencephalography (EEG) in the 1920s and
	made the first electrocorticogram in a human. Since then, EEG has been used to
	measure electrical activities generated by neuronal action potentials
	\cite{zhang01,henry06}.
	
	Mathematical models have been considered to reproduce neuronal action
	potentials. In 1907, Lapicque \cite{lapicque07} developed a neuron model based
	on an electric circuit composed of a capacitor and resistor in parallel
	\cite{burkitt06}. Hodgkin and Huxley \cite{hodgkin52} proposed in 1952 a model
	containing terms related to different types of ion channels. In 2005, Brette
	and Gerstner \cite{brette05} introduced the adaptive exponential
	integrate-and-fire (AEIF) model. AEIF is a model that has an exponential spiking
	mechanism combined with an adaptation equation \cite{naud08}. Neuronal networks
	composed of AEIF neurons have been used to study synchronous behaviours
	\cite{borges17}, self-sustai\-ned activity \cite{borges20}, and firing patterns
	\cite{santos19}. Shiau and Buhry \cite{shiau19} analysed interneuronal gamma
	oscillations in the hippocampus by means of AEIF neurons.
	
	The activity of isolated neurons can be affected by different types of noise.
	There are many sources of noise in the brain, such as from genetic processes,
	thermal noise, ionic channel fluctuations, and synaptic ev\-ents
	\cite{faisal08}. It was reported that noisy neurons are critical for learning
	\cite{engel15}. Synaptic noise is a source of randomness in the neuronal
	interspike intervals \cite{calvin68}. Brunel et al. \cite{brunel01} studied the
	effects of synaptic noise on the frequency response of neurons. They found that
	noise inputs can enhance high frequency responses. Gonz\'alez-Villar et al.
	\cite{villar17} observed that higher neuronal noise can be related to cognitive
	dyscognition in chronic pain syndromes. It was reported that L\'evy noise
	\cite{xu13,xu15,xu16} can induce stochastic resonance in the FitzHugh-Nagumo
	(FHN) neuron \cite{wang16}, as well as temporal and spatial coherence in
	coupled FHN neurons \cite{wang19}.
	
	Neuronal activities with noise have been modelled by means of stochastic
	differential equations \cite{wu17,feng18}. Various stoch\-astic neuron models
	were introduced, for instance the stochastic integrate-and-fire \cite{lansky06}
	and Fitz\-hugh-Nagumo \cite{tuckwell03} neuronal models. The sto\-chastic
	Hodgkin-Huxley model \cite{goldwyn11}, where the noise is incorporated in the
	ion channel, is able to simulate the behaviour of cerebellar granule cells
	\cite{saarinen06}. Braun et al. \cite{braun11} studied a Hodgkin-Huxley type
	model subjected to a current noise. They found transitions from single to
	grouping of spikes in response to temperature alterations. 
	
	In this work, we include noise in the AEIF model and study the effects on 
	neuronal firings. To do that, we compute the neuronal interspike intervals and
	calculate their coefficient of variation. The coefficient of variation can be
	used to measure the variability of the interspike intervals and to identify
	tonic spiking and burst patterns. By varying the noise amplitude, we show the
	existence of continuous and noncontinuous transitions in the neuronal interspike
	intervals. The noncontinuous transitions occur mainly for parameters close to
	the border between tonic spiking and burst activities of the neuron without
	noise.
	
	The paper is organized as follows. In Section $2$, we introduce the stochastic
	model that mimics neuronal activity and the diagnostic tools. Section $3$
	presents our results about the effects of noise on neuronal activities.
	Finally, we draw our conclusions in the last section.
	
	
	\section{Methodology}
	
	\subsection{Stochastic neuron model}
	
	The adaptive exponential integrate-and-fire (AEIF) mo\-del has been used to
	mimic neuronal tonic spiking and burst activities. In this work, We consider a
	stochastic adaptive exponential integrate-and-fire (SAEIF) model, where a
	sto\-chastic term, which corresponds to a noise current input, is added to the
	AEIF model. The SAEIF neuron model is given by
	\begin{eqnarray}
	\label{eqIF}
	\frac{dV}{dt} & = & \frac{1}{C_{\rm m}}\left[ f(V) -w+I \right]+
	\zeta\left(t\right), \\
	\frac{dw}{dt} & = & \frac{1}{\tau_m}\left[a(V-E_{\rm L})-w\right],
	\end{eqnarray}
	where $V$ and $w$ correspond to the membrane potential and the adaptation 
	current of a single neuron, respectively. $f(V)$ is a function defined as
	\begin{eqnarray}
	f(V)=-g_{\rm L}(V-E_{\rm L})+
	g_{\rm L} {\Delta}_{\rm T}\exp\left(\frac{V-V_{\rm T}}{{\Delta}_{\rm T}}\right).
	\end{eqnarray}
	In our simulations, we consider $C_{\rm m}=200$ pF (membrane capacitance),
	$g_{\rm L}=12$ nS (leak conductance), $E_{\rm L}$ $=-70$ mV (resting potential),
	$\Delta_{\rm T}=2$ mV (slope factor), $V_{\rm T}=-50$ mV (threshold potential),
	$\tau_w=300$ ms (adaptation time constant), and $a=2$ nS (level of subthreshold
	a\-daptation). A constant current $I$ equal to $500$ pA is injected to the
	neuron. We add a noise source $\zeta(t)$, where $\left<\zeta(t)\right>=0$ and
	$\left<\zeta(t)\zeta(t^{'})\right>=2D\delta(t-t^{'})$. $D$ is the noise
	amplitude and $\delta()$ is the Dirac delta function. We consider a refractory
	period of $1$ ms, namely a time in which the neuron can not fire. In the
	refractory period, the membrane voltage is maintained in the reset potential
	value $V_{\rm r}$. When the neuronal membrane potential is above a threshold
	($V>V_{\rm thres}$) \cite{naud08}, the variable states are updated according to
	\begin{eqnarray}
	V &\to & V_{\rm r}, \nonumber\\
	w &\to& w_{\rm r} = w + b ,
	\end{eqnarray} 
	where $V_{\rm r}$ and $b$ are the reset potential and the triggered adaptation
	addition, respectively. The set of e\-quations are solved by means of the
	Stochastic Runge-Kutta algorithm \cite{honeycutt92}.
	
	\subsection{Diagnostic tools} 
	
	Aiming to quantify the variability of the neuronal firings, we compute the 
	interspike intervals (ISI). The $m$-th ISI is defined as 
	\begin{equation}
	{\rm ISI}_m=t_{m+1}-t_{m},
	\end{equation}
	where $t_m$ is the time of the $m$-th neuronal fire. The coefficient of
	variation of ISI is then defined by
	\begin{equation}
	{\rm CV}=\frac{\sigma_{\rm ISI}}{\overline{\rm ISI}},
	\end{equation}
	where $\sigma_{\rm ISI}$ is the standard deviation and $\overline{\rm ISI}$ 
	the mean value of ISI. In the absence of noise ($D=0$), the neuron in Eq.
	\ref{eqIF} can exhibit tonic spiking and burst activities characterised by
	CV$<0.5$ and CV$\ge0.5$, respectively
	\cite{borges17,protachevicz18,protachevicz19}. The noise increases the
	irregularity of the firing times, and as a consequence, for intense noise
	amplitude, CV $,<\approx 1$ for neuronal tonic spiking and CV$>$1 for burst
	patterns \cite{gerstner14}. 
	
	In this work, to identify bifurcations in the range of the ISIs values, we
	compute the minimal ${\rm ISI}^{\rm min}$ and maximal ${\rm ISI}^{\rm max}$ value
	of the ISI in the parameter space $b\times V_{\rm r}$. For $D=0$, we identify a
	region where $({\rm ISI}^{\rm max}-{\rm ISI}^{\rm min})\le 0.5$ ms. When $D>0$,
	this difference increases and belongs to the same regions if the new ISI
	(${\rm ISI_{\rm new}}$) satisfies the inequality
	\begin{equation}
	\left[{\rm ISI}^{\rm min}_j-{\rm tol}(D)\right]\le{\rm ISI}_{\rm new}  
	\le\left[{\rm ISI}^{\rm max}_j+{\rm tol}(D)\right],
	\label{reltol}
	\end{equation}
	where the tolerance function (${\rm tol}$) is given by 
	\begin{equation}
	{\rm tol} (D)= 95\cdot D^{0.25}.
	\end{equation}
	
	\begin{figure}[ht]
		\centering
		\includegraphics[scale=0.075]{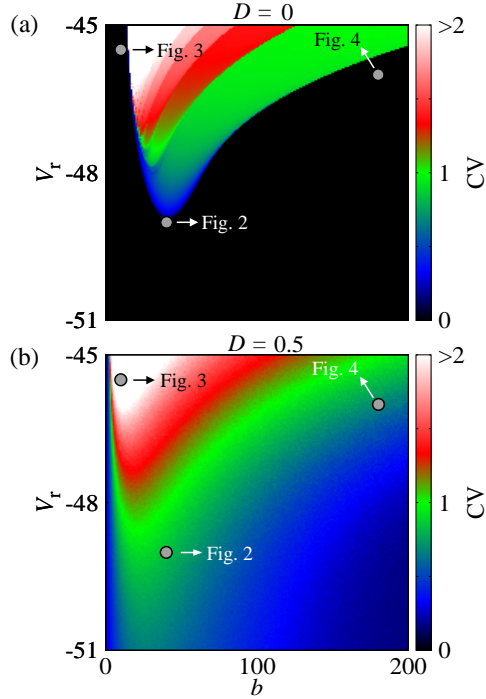}
		\caption{Parameter space $V_{\rm r}\times b$ for (a) $D=0$ and (b) $D=0.5$,
			where the colour bar corresponds to the CV values.}
		\label{fig1}
	\end{figure}
	
	When ${\rm ISI}_{\rm new}$ does not satisfy Eq. (\ref{reltol}), a noncontinuous
	transition occurs in the ISI, and as a consequence, $j$ new regions with
	different ISI ranges,
	\begin{equation}
	\left({\rm ISI}^{\rm max}_{\rm new}-{\rm ISI}^{\rm min}_{\rm new}\right)\le 40
	\;\;{\rm ms},
	\end{equation}
	appear. Regions larger than $40$ ms are not considered as discrete transitions,
	as well as regions with a small number of visits ($<10$) are not considered when
	identifying new regions.
	
	
	\section{Results}
	
	The coefficient of variation (CV) has been used to identify tonic spiking and
	burst patterns. Figure \ref{fig1} shows the CV values (colour bar) in the
	parameter space $V_{\rm r}\times b$ for (a) $D=0$ and (b) $D=0.5$. Increasing $D$
	from $0$ (Fig. \ref{fig1}(a)) to $0.5$ (Fig. \ref{fig1}(b)), increases the
	values of CV above $0$. The changes in CV when $D$ increases, indicate
	alterations in the dynamical behaviour. Due to this fact, we compute
	${\rm ISI}_m$ for different values of $V_{\rm r}$ and $b$ to understand the noise
	effect on neuronal dynamics.
	
	\begin{figure}[ht]
		\centering
		\includegraphics[scale=0.07]{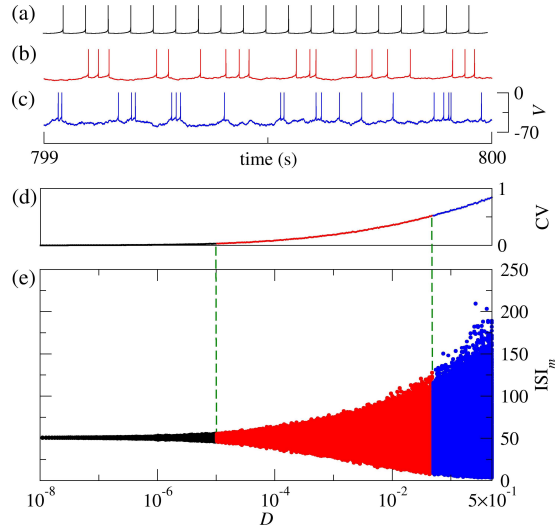}
		\caption{Temporal evolution of $V$ for (a) $D=0$, (b) $D=0.05$, and (c)
			$D=0.5$ for $V_{\rm r}=-49$ mV and $b=40$ pA. (d) CV and (e) ${\rm ISI}_m$ as a
			function of $D$.}
		\label{fig2}
	\end{figure}
	
	\begin{figure}[ht]
		\centering
		\includegraphics[scale=0.07]{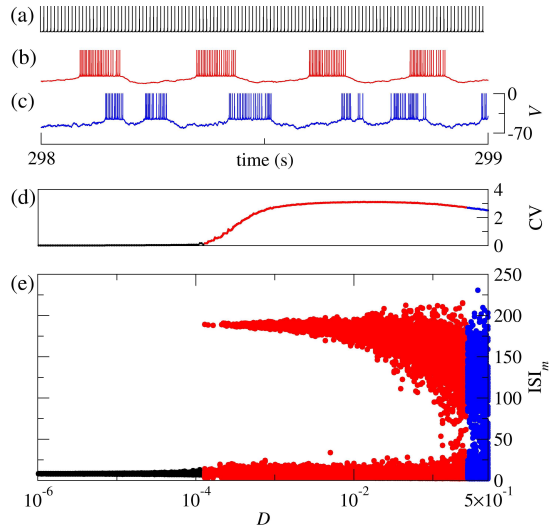}
		\caption{Temporal evolution of $V$ for (a) $D=0$, (b) $D=0.05$, and (c)
			$D=0.5$ for $V_{\rm r}=-45.5$ mV and $b=10$ pA. (d) CV and (e) ${\rm ISI}_m$ as
			a function of $D$.}
		\label{fig3}
	\end{figure}
	
	For $V_{\rm r}=-49$ mV, $b=40$ pA, and without noise ($D=0$), the neuron exhibits
	tonic spiking activities wi\-th an ISI value approximately equal to $50$ ms
	(see Fig. \ref{fig2}(a)). Depending on the noise amplitude, the neuronal pattern
	can display tonic spiking or burst activities, as shown in panels (b) and (c) in
	Fig. \ref{fig2}, respectively, and as a consequence, CV increases (Fig.
	\ref{fig2}(d)). In Fig. \ref{fig2}(e), we observe a continuous increase of the
	ISI range when the noise amplitude is increased from $10^{-8}$ to $0.5$. Figure
	\ref{fig3} displays a situation in which a noncontinuous change in the values
	of the ISI range occurs, where we consider $V_{\rm r}=-45.5$ mV and $b=10$ pA.
	The neuron exhibits tonic spiking activity for $D=0$ with ISI value
	approximately equal to $8$ ms  (Fig. \ref{fig3}(a)). The tonic spiking
	activities turn into bursts for $D=0.05$ (Fig. \ref{fig3}(b)) and $D=0.5$ (Fig.
	\ref{fig3}(c)), and as a result, the CV increases, as shown in Fig.
	\ref{fig3}(d). We find that for $D\approx 10^{-4}$, a new ISI range appears
	($\approx 190$ ms), namely the noise current induces a noncontinuous transition
	from $1$ to $2$ ISI ranges ($1\to 2$).
	
	\begin{figure}[ht]
		\centering
		\includegraphics[scale=0.07]{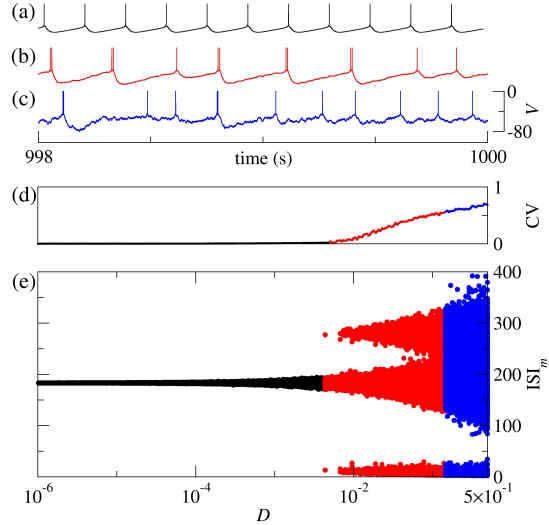}
		\caption{Temporal evolution of $V$ for (a) $D=0$, (b) $D=0.05$, and (c)
			$D=0.5$ for $V_{\rm r}=-46$ mV and $b=180$ pA. (d) CV and (e) ${\rm ISI}_m$ as
			a function of $D$.}
		\label{fig4}
	\end{figure}
	
	The unperturbed AEIF neuron exhibits tonic spiking activities for $V_{\rm r}=-46$
	mV and $b=180$ pA (Fig. \ref{fig4}(a)). Without noise, the interval time between
	tonic spiking activities is periodic, however a small noise amplitude is enough
	to alter the time interval of irregular behaviour, as shown in panels (b) and
	(c) in Fig. \ref{fig4}. Moreover, the noise also increases the values
	of CV (Fig. \ref{fig4}(d). The increase of CV begins when the $1\to 3$
	transition occurs (Fig. \ref{fig4}(e)).
	
	\begin{figure}[ht]
		\centering
		\includegraphics[scale=0.2]{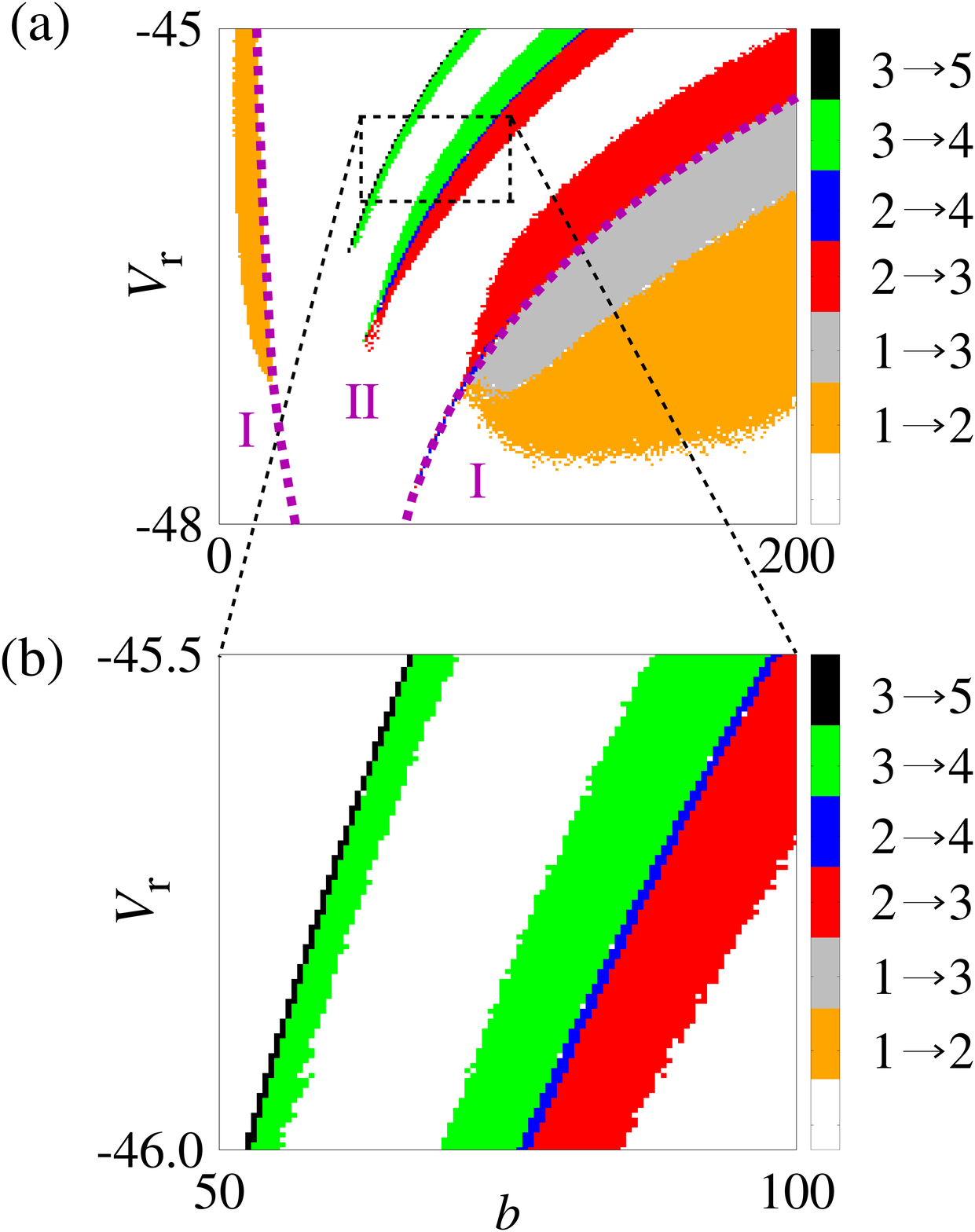}
		\caption{(a) $V_{\rm r}\times b$ showing the regions where continuous (white) and
			noncontinuous transitions of the ISI range appear: $1\to 2$ (orange),
			$1\to 3$ (gray), $2\to 3$ (red), $2\to 4$ (blue), $3\to 4$ (green), and
			$3\to 5$ (black). (b) Magnification of the parameter space in the intervals
			$V_{\rm r}=[-46,-45.5]$ mV and $b=[50,100]$ pA. Regions I and II correspond to
			tonic spiking and burst activities, respectively.}
		\label{fig5}
	\end{figure}
	
	In Fig. \ref{fig5}(a), we identify the regions in the parameter space where the
	continuous (white) and noncontinuous transitions (other colours) appear. In our
	simulations, we find the following noncontinuous transitions of the ISI
	range: $1\to 2$ (orange), $1\to 3$ (gray), $2\to 3$ (red), $2\to 4$ (blue),
	$3\to 4$ (green), and $3\to 5$ (black). Figure \ref{fig5}(b) displays the
	magnification of the parameter space in the intervals $V_{\rm r}=[-46,-45.5]$ mV
	and $b=[50,100]$ pA. The transition is identified through $50$ repetitions with
	an analysed time equal to $25$ s, where the transient time is set equal to $1$
	s.
	
	
	\section{Conclusions}
	
	Neurons are nerve cells responsible for receiving and transmitting information
	in the brain. Mathematical models have been developed and used to describe
	neuronal behaviour. The adaptive exponential integrate-and-fire (AEIF) model
	has been considered to mimic neuronal tonic spiking and burst activities. In
	this work, we include a stochastic term in the AEIF model to analyse the effect
	of noise in the activity of the neuron. The noise is inherent to neuronal
	activities and arise from several sources.
	
	We find that noise induces alteration in the interspike interval (ISI), and as
	a consequence, the neuronal activities can change from tonic spiking to burst
	activites and vice versa. We observe that bifurcations in the ISI range occur
	when the noise amplitude is increased. The bifurcations are noncontinuous
	transitions and depend not only on the noise amplitude, but also on the reset
	potential ($V_{\rm r}$) and the triggered adaptation addition ($b$). We find the
	noncontinuous bifurcations in the regions of the parameter space
	$V_{\rm r}\times b$ close to the border in which the unperturbed neuron changes
	from tonic spiking to burst patterns. In the parameter space with burst pattern,
	we also observe discrete bifurcations close to the border in which unperturbed
	neurons exhibit different number of fires per burst. Therefore, noise plays an
	important role in the neuronal behaviour related to tonic spiking and burst
	activities, as well as to ISI ranges.
	
	
	\section*{Acknowledgment}
	This study was financially supported by the following Brazilian government
	agencies: Fun\-da\c c\~ao Arauc\'aria, National Council for Scientific and
	Technological Development, Coordination for the Improvement of High\-er
	Education Personnel, and S\~ao Pa\-ulo Research Foundation (2015/07311-7,
	2017/18977-1, 2018/03211-6).
	


\begin{thebibliography}{99}
		\bibitem{caton75}
		R Caton, Brit. Med. J. {\bf 2} 278 (1875)
		\bibitem{coenen13}
		A Coenen and O Zayachkivska, Adv. Cogn. Psychol. {\bf 9} 216 (2013) 
		\bibitem{vaque99}
		T J La Vaque, J. Neurother. {\bf 3} 1 (1999)
		\bibitem{zhang01}
		L I Zhang and M-M Poo, Nat. Neurosci. {\bf 4} 1207 (2001)  
		\bibitem{henry06}
		J C Henry, Neurology {\bf 67} 2092 (2006)  
		\bibitem{lapicque07}
		L Lapicque, J. Physiol. Pathol. Gen. {\bf 9} 620 (1907)
		\bibitem{burkitt06}
		A N Burkitt, Biol. Cybern. {\bf 95} 1 (2006)
		\bibitem{hodgkin52}
		A L Hodgkin and A F Huxley, J. Physiol. {\bf 116} 449 (1952)
		\bibitem{brette05}
		R Brette and W Gerstner, J. Neurophysiol. {\bf 94} 3637 (2005)
		\bibitem{naud08}
		R Naud, N Marcille, C Clopath and W Gerstner, Biol. Cybern. {\bf 99} 335 (2008)
		\bibitem{borges17}
		F S Borges, P R Protachevicz, E L Lameu, R C Bonetti, K C Iarosz, I L
		Caldas, M S Baptista and A M Batista, Neural Netw. {\bf 90} 1 (2017)
		\bibitem{borges20}
		F S Borges, P R Protachevicz, R F O Pena, E L Lameu, G S V Higa, A H Kihara, F
		S Matias, C G Antonopoulos, R de Pasquale,  A C Roque, K C Iarosz, P Ji and  A
		M Batista, Physica A  {\bf 537} 122671 (2020)
		\bibitem{santos19}
		M S Santos, P R Protachevicz, K C Iarosz, I L Caldas, R L Viana, F S
		Borges, H P Ren, J D Szezech Jr, A M  Batista and C Grebogi, Chaos {\bf 29}
		043106 (2019)
		\bibitem{shiau19}
		L Shiau and L Buhry, Neurocomputing {\bf 331} 220 (2019)  
		\bibitem{faisal08}
		A A Faisal, L P J  Selen and D M Wolpert, Nat. Rev. Neurosci. {\bf 9} 292 (2008)
		\bibitem{engel15}
		T A Engel, W Chaisangmongkon, D J Freedman and X-J Wang, Nat. Commun. {\bf 6}
		6454 (2015)
		\bibitem{calvin68}
		W H Calvin and C F Stevens, J. Neurophysiol. {\bf 31} 574 (1968)
		\bibitem{brunel01}
		N Brunel, F S Chance, N Fourcaud and L F Abbott, Phys. Rev. Lett. {\bf 86} 2186
		(2001)
		\bibitem{villar17}
		A J Gonz\'alez-Villar, N Samartin-Veija, M Arias and M T Carrillo-de-la-Pe\~na,
		Sci. Rep. {\bf 7} 5841 (2017)
		\bibitem{xu13}
		Y Xu, J Feng, J J Li and H Q Zhang, Chaos {\bf 23} 013110 (2013)
		\bibitem{xu15}
		Y Xu, J Feng, W Xu and R C Gu, CMES {\bf 106} 309 (2015)
		\bibitem{xu16}
		Y Xu, Y G Li, H Zhang, X F Li and J Kurths, Sci. Rep. {\bf 6} 31505 (2016)
		\bibitem{wang16}
		W ZhanQing, X Y Yong and Y Hui, Sci. China Technol. Sc. {\bf 59} 371 (2016)
		\bibitem{wang19}
		Z Wang, Y Xu, Y Li and J Kurths, J. Stat. Mech. Theory Exp. {\bf 2019} 103501
		(2019)
		\bibitem{wu17}
		J Wu, Y Xu and J Ma, PLoS ONE {\bf 12} e0174330 (2017)
		\bibitem{feng18}
		J Feng, W Xu, Y Xu and X L Wang, Physica A {\bf 531} 121747 (2018)
		\bibitem{lansky06}
		P Lansky, P Sanda, J. Comput. Neurosci. {\bf 21} 211 (2006)
		\bibitem{tuckwell03}
		H C Tuckwell, R Rodriguez and F Y M Wan, Neural Comput. {\bf 15} 143 (2003)
		\bibitem{goldwyn11}
		J H Goldwyn, N S Imennov, M Famulare and E Shea-Brown, Phys. Rev. E {\bf 83}
		041908 (2011)
		\bibitem{saarinen06}
		A Saarinen, M-L Linne and O. Yli-Harja, Neurocomputing {\bf 69} 1091 (2006)
		\bibitem{braun11}
		H A Braun, J Schwabedal, M Dewald, C Finke, S Postnova, M T Huber, B Wollweber,
		H Schneider, M C Hirsch, K Voigt, U Feudel and F Moss, Chaos {\bf 21} 047509
		(2011)
		\bibitem{honeycutt92}
		R L Honeycutt, Phys. Rev. A {\bf 45} (1992) 600 (1992)
		\bibitem{protachevicz18}
		P R Protachevicz, R R Borges, A S Reis, F S Borges, K C Iarosz, E L Lameu,
		E E N Macau, R L Viana, I M Sokolov, F A S Ferrari, J Kurths, A M Batista,
		C-Y Lo, Y He and C-P Lin, Physiol. Meas. {\bf 39} 074006 (2018)
		\bibitem{protachevicz19}
		P R Protachevicz, F S Borges, E L Lameu, P Ji, K C Iarosz, A H Kihara, I L
		Caldas, J D Szezech Jr, M S Baptista, E E N Macau, C Antonopoulos, A M Batista
		and J Kurths, Front. Comput. Neurosc. {\bf 13} 1 (2019)
		\bibitem{gerstner14}
		W Gerstner, W N Kistler, R Naud and L Paninski, Neuronal Dynamics: From single
		neurons to networks and models of cognition, Cambridge University Press (2014)
	\end{thebibliography}
\end{document}